# Wireless Authentication System for Barcode Scanning Using Infrared Communication Technique

*M.S. Raheel, M. R. Asfi, M. Farooq-i-Azam, H. R. Shaukat, J. Shafqat
Department of Electrical Engineering
COMSATS Institute of Information Technology
Lahore, Pakistan
*msraheel@ciitlahore.edu.pk, fazam@ciitlahore.edu.pk*

***Abstract --*** *This paper presents the verified methodology for wireless authentication system using infrared barcode based scanner. An alternate approach of scanning a barcode using an infrared communication is implemented here, comprises of a card on which a barcode is printed is illuminated with an infrared beam. The reflections from the card are received by an infrared receiver and transmit them to a remote machine over wireless channel where the retrieve information is further processed and converted into the corresponding barcode digits which are further matched against the database stored in the server and the authenticity of the user is established. The system developed is cost-effective, efficient, provide double security and robust.*

***Keywords*** *– Infrared Communication, Barcodes, Wireless, VC++.*

## 1. INTRODUCTION

Traditionally, a variety of authentication systems have been used in various applications. Some of the very common systems include, password based authentication, smart cards, biometric systems and barcode cards. Password based authentication system is one of the oldest and simplest way of authenticating an individual. Some older and one of the first authentication systems were those of having barcode detection mechanism. Different standards of barcodes were developed to encrypt the information and interpreted when required. Standards, such as CODE 39, CODE 128, EAN-16, etc which are described in [1]. Those systems, still active today, work on the principles of laser based technology i.e. laser beam is used to decrypt the information written in form of barcode.

In 2000, a research evaluated the uses, characteristics, advantages and drawbacks of these systems on a comparative basis [2]. Image based authentication is a different type in which graphical characteristics of a person like shape of the hands or facial characteristics are analyzed and matched with the database of stored images for authenticating an individual as well as for detection of vehicles number plates [3]. As far as the voice based authentication is concerned, research has been conducted upon such a voice-coded system where the voice pattern is authenticated with the margins of change in the tone of the person, thus making the system more secure, this system has been deployed in the form of a standalone device as well as in form of a robot based surveillance system [4]. A secure authentication captured the markets and still is very effective today. This is called Smart Card based authentication. A magnetic strip containing the coded information is used for verifying a certain individual. Different security concerns and data integrity issues were addressed by the passage of time [5].

Biometric systems are being used in different scenarios which contain the academic attendance maintenance using fingerprints scanner, retinal scan so that relational databases can be made more secure and to avoid the use of extra storage channels in the same database. On the other hand, smart card systems are used in scenarios involving ATM transactions and also the attendance systems in examinations procedures [6-8].

Coming back towards barcode authentication, In addition to the above methodologies, barcode cards have also been used to authenticate users to a system. Barcode cards are read using an appropriate scanner. Traditionally, laser-based scanners have been used for the purpose as described before.

In this publication, the proposed architecture comprises of two main stages. First the user is asked a password and if the correct password is provided, only then the user can proceed to the next and more secure authentication stage based upon barcode. Barcode scanning and detection mechanism is based upon the fact that black color absorbs infra-red radiations. On the other hand white color reflects the radiations thus enabling to detect the printed barcode. The pulses received by the receiver will be transmitted to a personal computer for comparison of detected code with those stored in the database. The persons who are to be authenticated using this system are issued barcode cards with unique codes. Every barcode corresponds to a unique integer value. Barcode standard followed for the implementation is the widely used CODE 39 which also has immense commercial importance.

## 2. PROPOSED METHODOLOGY

The proposed barcode scanning and authentication system comprises of two major parts:
  i. a stand-alone device that scans a barcode card and then transmits the code to a remote system using a wireless link.

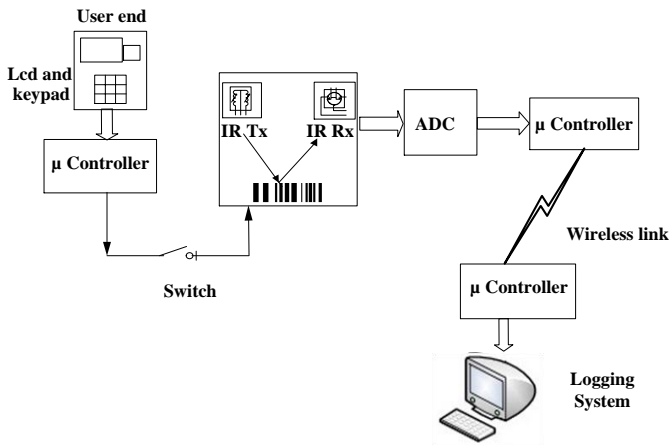

Fig. 1 Proposed Wireless Authentication System description.

ii. a remote processing and logging system that receives the code from the barcode scanner over a wireless link and then compares the received code to the set of allowed codes in the database.

The logging system (computer system) displays an appropriate message according to the result of the comparison. A flowchart of the proposed authentication system is shown in Figure 1 and a block diagram is shown in Figure 2.

The infra-red barcode scanning device is protected by a password. The password authentication is implemented using a keypad, LCD module and microcontrollers. Only after the user is authenticated by entering the password in the first stage, the scanning system allows the user to scan a barcode card in the second stage. If a user fails to enter correct password three times, the scanning device is turned off and can only be reset by a field engineer using a factory pre-set code. After successful password authentication, the card is inserted in a tray which is mounted on top of a DC motor. The motor slides the card in front of an infra-red transmitter using a constant speed so as to avoid scanning errors.

## 3. DESIGN DESCRIPTION

When the card is being scanned, the infra-red radiations transmitted by the infra-red transmitter strike the surface of the card. Black strips absorb the radiations and the white strips reflect them. The reflected radiations are received by an infra-red receiver which provides an output in the form of a voltage signal. This voltage signal is fed to an 8-bit parallel analog-to-digital converter (ADC) to generate a corresponding digital bit stream. The barcode in the form of digital bit stream is then further processed by an 8-bit microcontroller and sent to a wireless module connected to the microcontroller so as to transmit it to the remote system using amplitude shift keying (ASK) at 434 MHz. The data transmitted is raw data which just indicates presence or absence of black and white strips. It still does not represent the integer value of the barcode. The remote system receives the barcode using a corresponding 434 MHz ASK receiver which is controlled by a second microcontroller at the remote system. This receiving-side microcontroller sends the received data to the computer through its serial port using EIA-232.

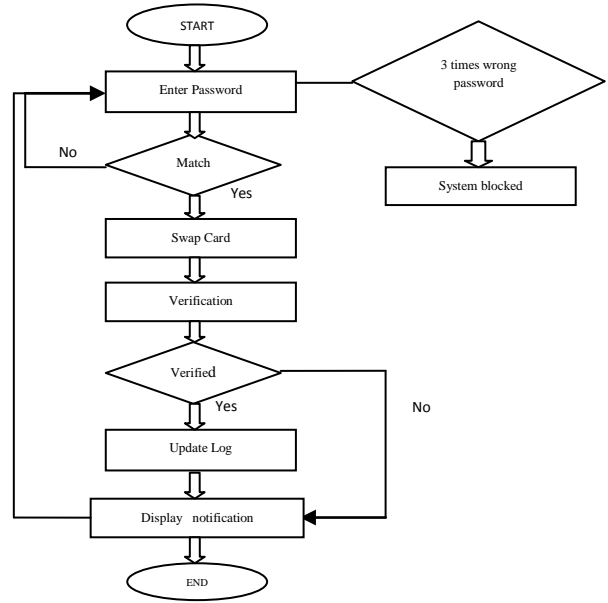

Fig. 2 Block Diagram.

An application program running on the computer receives the barcode from the serial port and further processes it performing error correction so as to discard redundant bits. It also converts the raw data which just represents the presence and absence of black and white strips into an integer value of the barcode according to barcode standard code 39. This value is then used to match the codes of authorized users stored in a database file, which is just a text file created for the purpose of this experiment. If the received barcode matches a barcode value in the database of authorized users, the name of the user to whom the barcode is issued and the time of authentication are logged in another log file which can later be used to perform checks such as which users entered the facility and at which times. The authentication result is also sent back to the scanning system which displays it on the LCD display for the user.

For each of the three microcontrollers, respective programs were developed in C language. Similarly, the software application running at the logging system and developed as part of this research does a major job in processing the received raw data, error detection and correction, and then conversion of the raw data to a corresponding integer barcode value.

## 4. EXPERIMENTAL SETUP

The core of the system comprises of the following hardware components:

i. Three 8-bit microcontrollers
ii. Infra-red transmitter and receiver
iii. Analog-to-digital converter (ADC)
iv. ASK wireless transmitter and receiver
v. Computer System, LCD and Keypad

On the scanning device, one microcontroller along with the keyboard and LCD is used for authentication of the user through a password. The user enters the password using the keyboard and is matched against the fixed value stored in the controller ROM. LCD is used to display the appropriate messages during all this process.

The second microcontroller is responsible for the actual scanning process. It is this microcontroller that turns off or on the infra-red transmitter and receives raw data generated by the infra-red receiver through an 8-bit parallel ADC [5]. The transmitter used for scanning is a NAND oscillator infra-red transmitter. The data generated after processing by the ADC is in the form of simple 1's or 0's. 1''s are generated when the scanning device sees a white bar and 0's are generated when the scanning device sees a black bar. The data is in the raw form as it does not as yet correspond to the integer value of the barcode. This raw data is further processed by the microcontroller and it transmits a character '5' for a byte of all 1's and transmits character '0' for a byte of all 0's. The wireless transmitter used is TWS-434 which uses amplitude shift keying to transmit the digital bit stream and operates at 434 MHz [10]. This particular frequency band does not require licensing to operate and is, therefore, widely used for short to medium range communication for special purposes.

The remote logging system houses an RWS-434 ASK receiver operating on the same band of 434 MHz. The output provided by the wireless receiver is a digital bit stream which is TTL compatible. This digital data is sent to the microcontroller interfaced to the wireless receiver. The microcontroller further transmits this data to the processing and logging computer system over EIA-232. This data is further processed by the software application running on the system and converts the data from 0's and 5's to an integer value corresponding to the actual barcode.

## 5. FUNCTION TESTING

After integrating the system, testing was done on the whole system and results were gathered. They were carefully analyzed. There are some functional requirements needed to be verified before proceeding to the results. Multiple tests were conducted for each critical component and their functionality was thoroughly verified. The summary of the tests and results is given in Table I and Table II.

## 6. RESULTS AND ANALYSIS

As part of our implementation of the system, different tests were performed at various stages of the development of the system. A major problem that we encountered was the detection and correction of error in the bit stream received through the wireless link at the remote logging system. At first, we tested our system over a wired channel operating at 9600 bps over EIA-232 between the scanning device and the remote logging system. The bit error rate thus encountered was of the order of 0.5%. Using wireless link, the error rate increased to 5%, which is an increase of 10% over that of wired link. The increased error rate expected in the wireless channel was due to a number of factors major being external noise and inter-symbol interference.

TABLE I
USER END HARDWARE TESTING TABLE

| Requirement Tested | CYCLE 1 | CYCLE 2 | FINAL STATUS | DETAILS |
|---|---|---|---|---|
| FR-02-001 | OK | Ok | Ok | Test of LCD and Keypad |
| FR-02-002 | Ok | Ok | Ok | Test of Password Protection System |
| FR-02-003 | Failed | Ok | Ok | Test of ADC free running mode |
| FR-02-004 | Ok | Ok | Ok | IR transmitter and Receiver test |
| | Failed | Ok | Ok | Test for alignment in the box |
| FR-02-005 | Ok | Ok | Ok | Serial communication test |
| | Failed | Ok | Ok | RF modules test |
| FR-02-006 | Ok | Ok | Ok | Barcode detection test |
| FR-02-007 | Ok | Ok | Ok | Test for FM transmitter and receiver |

TABLE II
SERVER END APPLICATION TESTING TABLE

| Requirement Tested | CYCLE 1 | CYCLE 2 | FINAL STATUS | DETAILS |
|---|---|---|---|---|
| FR-03-001 | Ok | Ok | Ok | Server end VC++ application |
| FR-03-002 | Failed | Ok | Ok | Real time reception of data |
| FR-03-003 | Ok | Ok | Ok | Filing of the results |

Though the problem could be resolved by introducing a data link layer for error detection and correction, we adopted the easier approach and wrote a dedicated error detection and correction routine in the software application running at the logging system. It is based upon a simple algorithm to convert any two zero's between 5's into 5 as well. This brought down our error rate low enough so as to approach that of wired interface.

The raw data obtained after error detection and correction was further processed to produce an array of 0s and 1s depending upon the number of continuous bits compared with the threshold calculated from the continuous bit stream. Bit streams having entries greater than that of calculated threshold were assigned the value 1 and those having number of entries less than threshold were assigned the value 0.

A 36-bit array containing the information about thick or thin bars of the scanned barcode in a form of 0's and 1's is obtained. Since IR reflections vary based upon the thickness of the bars in the printed barcode, we get different numbers of low and high voltages which are represented by 0's and 5's. When a variation from black to white or white to black occurs, the storage of bits is moved in the next index of an array.

The maximum and minimum numbers of array are calculated and a threshold is defined which then helps to produce a binary data array because barcode integers are represented as a combination of 0 and 1. Figure 5 shows barcode array after the described processing. The 36-bit array string is broken down into 4 parts each of 9 bits. Each 9-bit code represents a unique integer value. The 9-bit code is looked up in a table of values to yield the corresponding integer.

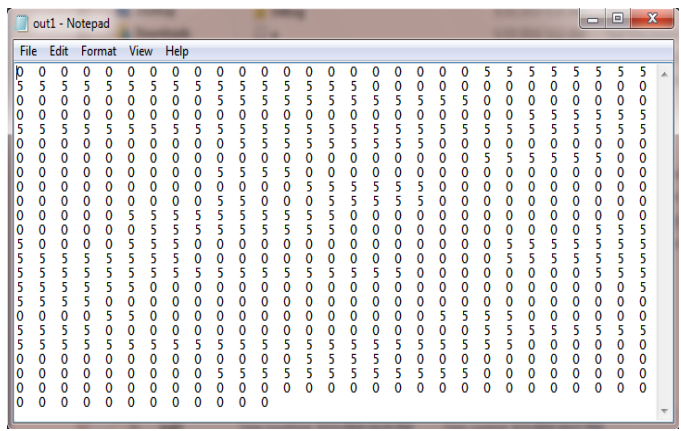

Fig. 4 Raw Data after error detection and correction.

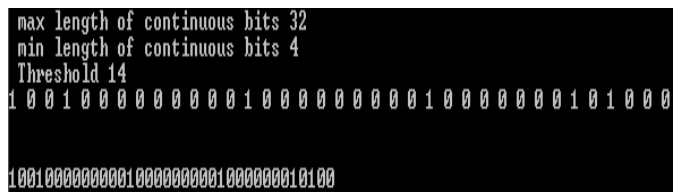

Fig. 5 Application showing Decoded Barcode Bit Stream.

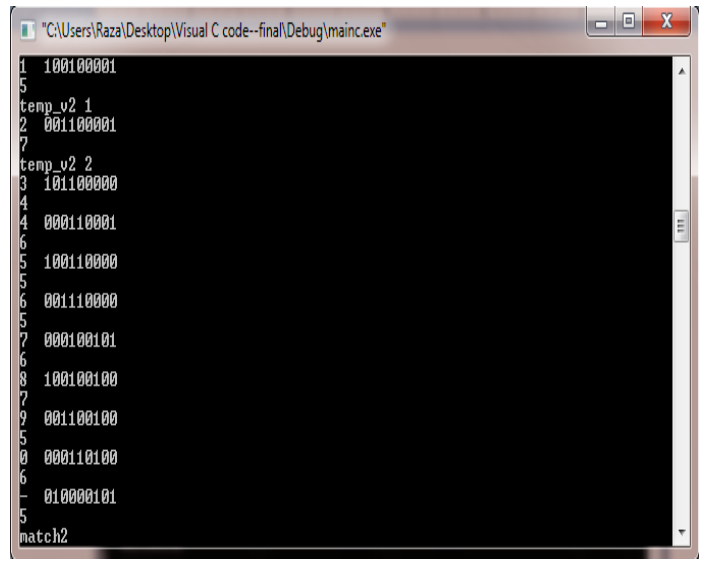

Fig. 6 Single integer match result.

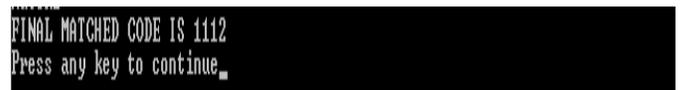

Fig. 7 Final matched barcode result.

In the Figure 6 below, the horizontal 9 bit strings are the code values along with the digit they represent, then a variable temp_v2, showing the bits matched against one integer. This variable temp_v2 then passes its value to another variable which only holds one value which is the highest value of the matched bits against an integer and finally the matched digit based upon the maximum number of bits matched out of 9 is displayed for each digit in database.

The above process of single integer match is performed four times to get the result of all four integers in the form of a single number which is then used to look into the database for the barcode match and log entries. A result of final matched barcode is shown in the Figure 7 below.

After the calculation of the barcode integer value, it is used, to authenticate the user by matching it against stored values in the database as described earlier. If the user is authenticated, an entry in the log file is made with the current date and time put next to the barcode integer value.

## 7. CONCLUSION

The development of an authentication system using infra-red barcode scanning which can be used as low-cost alternative to laser scanning has been presented in this paper. An infrared communication approach is used and the proposed system has tested and various results, which were obtained as part of this research, have been presented. In addition, we have also demonstrated that a scanning device needs not to be a standalone. It can rather communicate with a central system where the information is actually processed and stored. To make the system more robust, a data link layer can be introduced in the microcontroller software which will act as a driver for the wireless transmitter and receiver. The data link layer can be used for error detection and correction. Infrared scanners are of high cost and may not be applicable in environments where use of laser is not permitted

## Acknowledgement

The authors would like to acknowledge the respective department for providing adequate resources and environment for research.

## REFERENCES

[1] Harry E.Burke, "Automating Management Information Systems: Barcode engineering and Implementation", Van Nostrand Reinhold, ISBN 0-442-20712-3, (1990)

[2] Zwiesele, A, Munde A, Busch, C, Daum, H. "BioIS study: Comparative study of biometric identification systems" Proceedings of IEEE 34th Annual International Carnahan Conference on Security Technology, ISBN 0-7803-5965-8, pp.60-63, (2000)

[3] Srinath Akula, Veerabhadram Devisetty, "Image Based Registration and Authentication System", In proceedings of Midwest Instruction and Computer Composium, (2004)

[4] Heungkyu Lee, Hanseok Ko; "Voice code verification system using competing models for user entrance authentication", International Conference on Consumer Electronics, ISBN 0-7803-8838-0, January (2005)

[5] M. Hong, H.Guo, B.Luo, "Security Design for Multi-Service Smart Card Systems", Second International Conference on Future Generation Communication and Networking, ISBN 978-0-7695-3431-2, pp.299-304, (2008)

[6] T. Nawaz, S. Pervaiz, A. Korrani, A. Din, "Development of Academic Attendance Monitoring System Using Fingerprint Identification", International Journal of Computer Science and Network Security, Pakistan, Vol. 9, No. 5, May (2009)

[7] R. Mahajan, T. Gupta, S.Mahajan, and N. Bawa, "Retina as Authentication Tool for Covert Channel Problem", World Academy of Science, Engineering and Technology, Issue 56, pp 153-158, (2009)

[8] M. I. Moksin,N.M.Yasin , "The Implementation of Wireless Student Attendance System in an Examination Procedure", International Association of Computer Science and Information Technology - Spring Conference, ISBN 978-0-7695-3653-8, April (2009)